\documentstyle[pra,aps]{revtex}
\tighten%
\draft
\twocolumn
\begin{document}%
\title{Bell Theorem for Nonclassical Part of Quantum Teleportation
Process}
\author{Marek \.Zukowski \\
{\protect\small\sl Instytut Fizyki Teoretycznej i Astrofizyki,
Uniwersytet Gdanski, PL-80-952 Poland}}
\date{\today}
\maketitle
\begin{abstract}
The quantum teleportation process is composed of a joint
measurement performed upon two subsystems A and B (uncorrelated),
followed by a unitary transformation (parameters of which depend
on the outcome of the measurement) performed upon a third
subsystem C (EPR correlated with system B). The information about
the outcome of the measurement is transferred by classical means.
The measurement performed upon the systems A and B collapses their
joint wavefunction into one of the four {\it entangled} Bell
states. It is shown here that this measurement process plus a
possible measurement on the third subsystem (with classical
channel switched off - no additional unitary transformation
performed) cannot be described by a local realistic theory.

\end{abstract}

\vfill


{\it Quantum teleportation} \cite{BBCJPW} is the operational protocol
which enables one to transfer
 the
quantum state of one system, say $A$, to another quantum system, $C$.
  The transfer can be obtained
 by performing a joint (`Bell-state') measurement on $A$ and a third
system $B$, originally EPR entangled with $C$, and then
unitarily transforming $C$ according to the outcome of
this measurement.  Teleportation separates the
complete information in $A$ into two parts: a {\it classical}
part carried by the outcome $c$ of the joint measurement on
$A$ and $B$, and a {\it nonclassical} part carried by the
prior entanglement between $B$ and $C$.
 No
cloning of quantum information takes place; the
input state of $A$ is destroyed  before it is re-created
at $C$.

Teleportation is strongly related conceptually and experimentally to
other effects, like interferometric
tests of against local realism
involving independent sources of particles
~\cite{YS}, especially entanglement swapping ~\cite{{ZZHE},{ZZW}}.  In
entanglement swapping, the particle $A$ of the quantum
teleportation protocol
 is originally entangled with some particle $D$.
If, like in the case of
quantum teleportation,
a full Bell state measurement is performed on $A$ and $B$, and  depending
on the outcome, after a classical transfer of information,
a suitable unitary transformation is performed upon $C$,
the particle $D$ and $C$ are in an entangled state.
Thus, under such protocol, entanglement
swapping  can be interpreted as teleportation of entanglement
(from $A$ to $C$). The final state of $D$ and $C$
can be used in an experiment in which Bell inequalities are violated.
 The measurement acts on $D$ and $C$ can easily satisfy
the necessary requirement for a Bell inequality test, namely that
of spatial separation. Further, since the classical information on
the outcome of the Bell measurement is needed on one side only (in
the present example, in the vicinity of particle $C$),
 one may arrange the experiment in such a way
that no classical information on the result of the Bell-measurement
upon $A$ and $B$ can reach $D$ before the local measurement on $D$,
in the Bell inequality test, is done.
In such a case, the teleportation process can be treated
as just a more involved scheme of the preparation of the entangled state
of $D$
and $C$. This strongly suggests that there must be at least an
element in the teleportation procedure which defies local and realistic
interpretation.

The problem of the link of lack of link between the violations of
local realism  and the teleportation process has been addressed by
many authors (for  papers opening this discussion see
\cite{SANDU}). In this work the following aspect of the problem
will be discussed. As it was mentioned before, the teleportation
process has its quantum and classical part. The classical part
involves communication via standard classical methods, and thus
cannot be suspected of adding anything interesting to the relation
of the teleportation process with the Bell theorem (except for the
case of entanglement swapping, as discussed above). Even worse,
the classical transfer of information from the
Bell-state-measuring station (operated by Alice) to the particle
$C$ makes it possible that  measurements upon $C$ (after the full
teleportation protocol) can be  causally linked with the events at
Alice's apparatus. Thus a Bell type analysis is absolutely
excluded.
 Nevertheless,
as it will be argued below, the quantum part of the process cannot be
described by a local realistic formalism.

First one should define
 what is meant here
by the quantum part of the process. One can employ the simplest form
of amputation:
the classical information link between Alice and Bob is cut.
However, both parties are still allowed to perform the usual laboratory tasks
for an experiment towards verification of the actuality of
the teleportation process \cite{EXPVER}. Namely, Alice herself (or, for
purists, this can be done by
her friend Cecil)
can prepare the particle $A$ in any pure state, and subsequently
she can make a Bell measurement on $A$ and $B$. Bob, not knowing the
result  at Alice's side, nor the original state of $A$,
instead of being totally idle, performs on particle $C$ a
measurement of a (generally randomly chosen)
yes-no observable.

The formal description of the above runs as follows
(as in the classic case we assume all particles involved
to be two-state systems).
    The initial three particle state is:
\begin{eqnarray}
& (\sin{\beta}|A1\rangle + \cos{\beta}e^{i\phi}|A2\rangle)&\\
\nonumber
&\times\sqrt{\frac{1}{2}}(|B1\rangle|C1\rangle+|B2\rangle|C2\rangle),&
\end{eqnarray}
where $Ai$, $Bi$, $Ci$ denote the states of the three subsystems (the
letter stands
for the subsystem (particle) and $i=1,2$ is the index of two
orthogonal states). The parameters $\phi$ and $\beta$ are determined
by the state preparation procedure of Cecil.

Now Alice performs a measurement which collapses the $A-B$ system in to
the four Bell states:
\begin{equation}
\sqrt{\frac{1}{2}} (|B1\rangle|A1\rangle+|B2\rangle|A2\rangle) =
{\bf |00\rangle}, \label{BELL1}
\end{equation}
\begin{equation}
\sqrt{\frac{1}{2}} (|B1\rangle|A2\rangle+|B2\rangle|A1\rangle) =
{\bf |01\rangle},
\end{equation}
\begin{equation}
\sqrt{\frac{1}{2}} (|B1\rangle|A2\rangle-|B2\rangle|A1\rangle) =
{\bf |10\rangle},
\end{equation}
\begin{equation}
\sqrt{\frac{1}{2}} (|B1\rangle|A1\rangle-|B2\rangle|A2\rangle) =
{\bf|11\rangle}. \label{BELL4}
\end{equation}
Please note, that the names of the states introduced above are
binary expansions of $0,1,2,3$. They could constitute the possible
content of the classical messages of Alice to Bob, informing him
about the results obtained at the Bell-state analyzer (however,
this link is cut). The Alice's measurement projects  the particle
C into certain four states. Bob, cut off from Alice,  in
desperation performs an  experiment of a dychotomic (yes-no)
nature which results in the projections  into the two following
orthogonal states:
\begin{equation}
\cos{\beta'}|C1\rangle + \sin{\beta'} exp(i\phi')|C2\rangle =
|0\rangle
\end{equation}
and
\begin{equation}
-\sin{\beta'}|C1\rangle + \cos{\beta'} exp(i\phi')|C2\rangle =
|1\rangle
\end{equation}
The probabilities of all possible eight global results (2 results of
Bob times 4 results of Alice), are

\begin{eqnarray}
&P(00,0) = 1/4 - P(00,1)&\\ \nonumber &= \frac{1}{8}\big(1 -
\cos{2\beta} \cos{2\beta'} + \sin{2\beta}
\sin{2\beta'}\cos{(\phi-\phi')}\big),&
\end{eqnarray}

\begin{eqnarray}
&P(01,0) = 1/4 - P(01,1)&\\ \nonumber &= \frac{1}{8}\big(1 +
\cos{2\beta} \cos{2\beta'} + \sin{2\beta}
\sin{2\beta'}\cos{(\phi+\phi')}\big).&
\end{eqnarray}

\begin{eqnarray}
&P(10,0) = 1/4 - P(10,1)& \\  \nonumber &= \frac{1}{8}\big(1
+\cos{2\beta} \cos{2\beta'} - \sin{2\beta}
\sin{2\beta'}\cos{(\phi+\phi')}\big),&
\end{eqnarray}
\begin{eqnarray}
&P(11,0) = 1/4 - P(11,1)&\\ \nonumber & = \frac{1}{8}\big(1 -
\cos{2\beta} \cos{2\beta'} - \sin{2\beta}
\sin{2\beta'}\cos{(\phi-\phi')}\big).&
\end{eqnarray}

Let us assign to the four possible results of Alice's measurement,
$c=00,01,10$ or $11$, four  two-dimensional vectors (for some
other non-conventional value assignments for experimental results
see \cite{ZZH})
\begin{eqnarray}
\vec{A}(00)=(-1,-1),& \vec{A}(01)=(-1,1),& \nonumber \\
 \vec{A}(10)=(1,-1),& \vec{A}(11)=(1,1).&
\label{VECTORS}
\end{eqnarray}
The link between the vectors and the binary numbers is obvious.
The digit $0$ has been replaced by $-1$ because this trick makes the
subsequent
 derivation of a Bell inequality much easier.
Please note, that this procedure differs from the usual one (i.e.
assignment of certain real numbers, "eigenvalues", to certain
projectors) by the fact that we ascribe to the projectors more
complicated objects. The results of Bob's measurements, $i=0$ or
$1$, will be described in a similar fashion, namely by ascribing
numbers $I_B(0)=-1$ and $I_B(1)=+1$.

To simplify the description of the global measurement results one
can introduce a suitably defined correlation function.  Let us
consider such a function as the average of products of the results
on each side (here, vectors times numbers). E.g. the result
(00,0), i.e. a detection of the first Bell state, $00$, by Alice
and simultaneous detection of state $0$ by Bob, can be ascribed
$-1(-1,-1)=(1,1)$, etc. With such definitions of the values
assigned to the possible pairs of the outcomes the correlation
function
\begin{equation}
E(\beta, \phi; \beta',\phi') =
\sum_{c=00}^{11}\sum_{i=0,1}P(c,i)I_B(i)\vec{A}(c) \label{corr}
\end{equation}
acquires a form of a two dimensional vector and for the explicit
form of the quantum prediction
reads:
\begin{eqnarray}
   &E(\beta, \phi; \beta',\phi')_{QM}  &\\ \nonumber
&=\sin{2\beta}\sin{2\beta'}\left(\cos{\phi}\cos{\phi}',
\sin{\phi}\sin{\phi}'\right).&
\end{eqnarray}

Let us now simplify the problem a bit. Assume that Alice prepares
states of $A$ with
$\beta=45^o$, and Bob fixes his apparatus at $\beta'= 45^o$
too. The
correlation function is then simplified to:
\begin{eqnarray}
   &E(\phi;\phi')_{QM}  &\\ \nonumber
&=\left(\cos{\phi}\cos{\phi}',
\sin{\phi}\sin{\phi}'\right).&
\end{eqnarray}
It will be shown that  this correlation function
cannot be modeled by local hidden variable theories.

Imagine that a hidden variable $\lambda$ specifies the future results
of the experiments of Alice and Bob.
The product of such
predictions reads
\begin{equation}
I_B(\phi',\lambda)
\vec{A}(\phi,\lambda)
\label{hv}
\end{equation}
where, $I_B(\phi',\lambda) = \pm 1$ is the local hidden variable
(LHV) prediction for the result of the measurement by Bob  (for
the given value of the hidden parameter $\lambda$, and the local
observable defined by $\phi'$) and the vector
$\vec{A}(\phi,\lambda)$, which is the LHV prediction for the
Alice's result, depends on $\lambda$ and $\phi$, and takes one of
the four values (\ref{VECTORS}). The local hidden variable
prediction for the correlation function is an average of
(\ref{hv}) over a certain (properly normalized) distribution
$\rho(\lambda)$, namely
\begin{eqnarray}
   &E(\phi;\phi')_{LHV}  &\\ \nonumber
&=\int d\lambda \rho(\lambda)I_B(\phi',\lambda)
\vec{A}(\phi,\lambda).&
\end{eqnarray}
Now let us assume that Alice can set the values of the phase $\phi$ which
prepares the
state of particle $A$ at 0 or 90 degrees, whereas Bob can play with
$\phi'$ at -45 and +45 degrees.

To show that $E(\phi;\phi')_{QM}$ cannot be modeled by
$E(\phi;\phi')_{LHV}$  the  geometric   approach of \cite{ZUK} will be
used.
 It is
based on the following simple observation.
Assume that one knows the components of a certain
vector  $\bf q$ (the {\it known} vector) belonging to some vector space,
whereas about a second
vector $\bf h$ (the {\it test} vector) one is only
able to establish that its scalar product with $q$ satisfies
the inequality $\langle \bf h|\bf q \rangle < ||\bf q||^2$. The immediate
implication
is that these two vectors cannot be equal:
$\bf q\neq h$.

To form a vector for such an argument, one can
take the values of the quantum correlation function at the
$2\times2=4$ pairs of the possible settings of the macroscopic
parameters controlled by Alice and Bob $(\phi, \phi')$.
In this way
a "super-vector" $\bf V^{QM}$ is built.
The first component of the super-vector, for the settings $(0^o,
-45^o)$, reads:
\begin{equation}\vec{V}^{QM}_1=E(0;-45)_{QM}=
\left(
\sqrt{1/2}, 0\right)
,
\end{equation}
the second for $(0,45)$
\begin{equation}\vec{V}^{QM}_2=E(0;45)_{QM}=
\left(
\sqrt{1/2}, 0\right)
,
\end{equation}
the third one at $(90, -45)$
\begin{equation}\vec{V}^{QM}_3=E(90;-45)_{QM}=
\left(0,
-\sqrt{1/2}\right)
,
\end{equation}
the fourth $(90,45)$
\begin{equation}\vec{V}^{QM}_4=E(90;45)_{QM}=
\left(0,
\sqrt{1/2}\right)
.
\end{equation}

The square of the norm of such a super vector, denoted by
$\parallel\bf{V^{QM}}\parallel^2$, can be defined as the sum of
the squares of the norms of all the components, where the square
of the norm of a component is in turn the sum of the squares of
its two components. Therefore one has
\begin{equation}
\parallel {\bf
V^{QM}\parallel^2}=\sum_{i=1}^4|\vec{V}^{QM}_i|^2
= 2.
\end{equation}
We shall now estimate the scalar product of the quantum super-vector
with analogous super-vector $\bf V^{HV}$ which has the
structure characteristic for (deterministic) local hidden variables. Of
course, the aforementioned scalar product is defined in a way compatible
with the norm (i.e. it is a sum of the products
of the respective components, and the product of two components is again the
sum of the products of the respective  elements of the
components):
\begin{equation}
\big({\bf
V^{QM}, V^{LHV}}\big)=\sum_{i=1}^4\vec{V}^{QM}_i\cdot\vec{V}^{LHV}_i,
\end{equation}
with $\vec{V}^{LHV}_i$ being equal to the value of $E(\phi,\phi')_{LHV}$
for appropriate pairs of settings.
As it is usual in the proofs of the Bell theorem, it is  better
first to consider the hidden variable prediction for a single specified
$\lambda$ and
only later average this over the distribution of the hidden variables.

Thus, what we should do \cite{ZUK} is to estimate the scalar product of the
super-vector
constructed out of hidden-variable predictions for the specified
$\lambda$ with the quantum supervector (defined above).
The hidden variable super-vector  for a specific $\lambda$, which will
be denoted by $\bf{ H} (\lambda)$, has the following components:
\begin{equation}
\vec{ H} (\lambda)_1= I_B(-45,\lambda)
\big(A(0,\lambda)_1,A(0,\lambda)_2\big),
\end{equation}
\begin{equation}
\vec{ H} (\lambda)_2=I_B(45,\lambda)
\big(A(0,\lambda)_1,A(0,\lambda)_2\big),
\end{equation}
\begin{equation}
\vec{ H} (\lambda)_3=I_B(-45,\lambda)
\big(A(90,\lambda)_1,A(90,\lambda)_2\big),
\end{equation}
\begin{equation}
\vec{ H} (\lambda)_4=I_B(45,\lambda)
\big(A(90,\lambda)_1,A(90,\lambda)_2\big).
\end{equation}

For the scalar product $\big({\bf V^{QM}, H}(\lambda)\big)$, since
$I_B(\phi',\lambda)=\pm1$ and $A(\phi,\lambda)_i=\pm1$, one gets:
\begin{eqnarray}
&-2\sqrt{1/2} \leq \big({\bf V^{QM}, H}(\lambda)\big)&\\ \nonumber
&=\sqrt{1/2}\big(A(0,\lambda)_1[I_B(-45,\lambda) +
I_B(45,\lambda)]&
\\ \nonumber
 &+A(90,\lambda)_2[I_B(45,\lambda)
-I_B(-45,\lambda)]\big)\leq2\sqrt{1/2} .&
\end{eqnarray}
Thus if one now averages this inequality over the distribution of the
hidden variables $\rho(\lambda)$, the following relation emerges
\begin{eqnarray}
&-\sqrt{2}\leq ({\bf V^{QM},\bf V^{LHV}}) \leq  \sqrt{2} <
\parallel{\bf V^{QM}}\parallel^2= 2.& \label{Bell}
\end{eqnarray}
This implies simply that
$
\bf V^{LHV}\neq\bf V^{QM},
$
which means in turn nothing else than that {\it no local hidden variable
correlation function can reproduce the quantum prediction} (i.e. we
have a Bell theorem for the process). Please note that the appropriate Bell
inequality
is given here by the first two inequalities in (\ref{Bell}).

This method can still be expanded to cover much more settings
of the variables, here only the simplest case was presented.
It is an interesting fact that needs further investigation,
that the Bell inequality presented here is violated by the same factor
$\sqrt{2}$ as the CHSH inequality for the usual Bell theorem
involving a pair of particles in a maximally entangled state.
This may imply that the quantum component of the teleportation
process cannot be described in a local and realistic way
as long as the initial state of $B$ and $C$ neither admits such
models.

The present result also explains why the current local hidden variable
model explaining the low detection efficiency teleportation \cite{RAMON}
cannot be extended into high efficiency case. Simply, had this been possible,
such model would constitute a LHV model of the process considered here,
what by (\ref{Bell}) is impossible.
Also, for the same reason, considerations with toy-models
 like those in \cite{LUCIEN}
cannot be extended in such a way that they can fully reproduce the
quantum teleportation process. Nevertheless, the interesting conclusions
reached in \cite{GISIN}, namely that one can
model the teleportation process with  specific
local hidden variables {\it} and classical communication channel,
requiring the transfer of on average 2.19 bits, are not in a
disagreement with
the present result.

The inequality (\ref{Bell}) can  serve as the Bell-type inequality
for the experiment of Boschi {\it et al} \cite{BOSCHI}. In this
experiment the system A was replaced by the polarization degree of
freedom of one of the photons of the EPR entangled pair (the pair
was entangled in linear momentum  directions). In this way
measurement discriminating between the four correlated states of
polarization and momentum direction of a {\it single} photon,
which are formally equivalent to (\ref{BELL1}-\ref{BELL4}), can be
performed with standard quantum interferometric techniques. Thus
all observables involved in the present scheme found their
representation in the experiment. However, due to the angles
chosen in the experiment, one cannot directly apply the inequality
(\ref{Bell}). Nevertheless the obtained, very high, visibility of
the two-particle fringes, is well above the threshold ($71\%$)
indicated by (\ref{Bell}). This indirectly rules out a LHV model
for the experiment (of course, provided one accepts the fair
sampling assumption).

In the teleportation experiment involving all three particles
(with A emitted independently of the emission of the EPR pair $B$
and $C$) \cite{BOUWMEESTER},  due to fundamental technical
limitations, one currently cannot distinguish between all four
states (\ref{BELL1}-\ref{BELL4}), and therefore the inequality
cannot be applied. However, the extension of the experiment to the
teleportation of entanglement, i.e. entanglement swapping process
\cite{PAN}, \cite{BOUW2}, results in entangling previously
independent photons, on which in turn a Bell-type experiment can
be performed. Such experiment is possible on a subensemble of
events for which only one of the states (\ref{BELL1}-\ref{BELL4})
 was measured, i.e. entanglement swapping does not need
a full Bell-state measurement to be indescribable  by local
realistic theories. Unfortunately, the visibility in \cite{PAN}
was around $65\%$, i.e. within a zone for which one can build
explicit LHV models \cite{ZKBL}. Thus, higher visibility
realization of entanglement swapping
 would constitute an important fact in the empirical knowledge
on the nature of quantum teleportation.

The presented results cannot be applied directly to the
teleportation  experiment  involving continuous variables
of \cite{FURUSAWA}. However one can speculate
that the recent result of ref. \cite{BANASZEK}, concerning
the Bell theorem for the original EPR state, may, after suitable
extensions, lead to the same conclusions.

Support of the University of Gda\'nsk Grants No. BW-5400-5-0062-5
(1995),  BW/5400-5-0264-9 (1999) and of
 the Austrian-Polish Scientific Collaboration Program {\it
Quantum Information and Quantum Communication II} Nr. 11/98b is
acknowledged. The author wishes to thank for discussions Anton
Zeilinger, Harald Weinfurter   and the Horodecki Family.


\begin{thebibliography}{9}
\bibitem{EXPVER}
In an experiment verifying the process of teleportation one
 knows the state of $A$. However, such knowledge is
not required for a working quantum teleportation channel (this is
one of the most appealing features of the process).
\end{thebibliography}
\end{document}